\documentstyle[12pt, sprocl]{article}
\begin{document}
\title
{Multifractals of Normalized First Passage Time 
in Sierpinski Gasket }
\author
{ Kyungsik Kim and J. S. Choi}
\address
{Department of Physics, Pukyong National University, Pusan 608-737, 
Korea}
\author
{ Y.S. Kong$\dagger$ }
\address
{School of Ocean Engineering, Pukyong National University, Pusan 
608-737, Korea}
\maketitle\abstracts 
{  The multifractal behavior of the normalized first passage time is investigated 
on the two dimensional Sierpinski gasket with both
absorbing and reflecting barriers. 
The normalized first passage time for Sinai model and the logistic model to arrive at the
absorbing barrier after starting from an arbitrary site,
especially obtained by the calculation via the Monte Carlo simulation,
is discussed numerically.  
The generalized dimension and the spectrum  are also estimated from the
distribution of the normalized first passage time, and compared with
the results on the finitely square lattice.  }
\vspace{1.5cm}
Key words : normalied first passage time, generalized dimension, \hspace*{2cm}
logistc map, Sierpinski gasket, multifractal\\ 
\hspace*{11pt\hrulefill}\\
E-mail: kskim@dolphin.pknu.ac.kr, jschoi@dolphin.pknu.ac.kr\\
$\dagger$E-mail: yskong@dolphin.pknu.ac.kr\\
\section{Introduction}
Recently a lot of interest has been concentrated  on the behavior of the 
disorder system in a variety of contexts in condensed matter physics\cite{Vic1}-\cite{Hau4}.  
In the transport process one of the important subjects for the  regular and 
disorder systems is the random walk theory, and this theory 
has extensively been developed to 
the continuous-time random walk 
theory which is essentially described both by the transition probability 
dependent of the length
between steps and by the distribution of the pausing times \cite{Hau4}-\cite{Mon7}.
Until quite recent the transport phenomena for the motion of 
a particle have been largely
extended to the reaction kinetics \cite{Hau4,Mon71},
and the strange kinetics\cite{Mon72}.\\  
\indent
One of the well-known interesting issues in the theory of the
random walk  is the mean first passage time that is defined by 
the average time arriving at the
absorbing barrier for first time after a particle initially starts from an arbitrary
site. In particular, this problem  has extensively been
studied in the Sinai model which is discussed on the random barrier model
such as the absorbing and reflecting barriers.  
In previous work the Sinai model with asymmetric
transition probability also was studied for the mean and mean square
displacements dependent anomalously on time \cite{Sin2}-\cite{Mon6}. 
Moreover, the transport process for the mean first passage time has already been 
addressed by Noskowicz et
al.\cite{Nos8} who have obtained the upper and lower bounds using the recursion - relation
procedure \cite{Mur81}-\cite{Mur10}.  In one dimensional lattice with the periodic
boundary condition, Kozak et al.\cite{Koz11}  
treated long ago with the first passage time of a particle for the case of the
transition probability as a chaotic orbits of logistic map \cite{Pen12},
and obtained the statistical value of the mean first passage time 
with the anomalous exponent. \\
\indent
On the other hand, since the multifractal characteristics is directly related to 
analyze the statistical property of the normalized first passage time,
the multifractal for the mean first passage time 
has been studied intensively in connection
with the random walk problem in the one dimensional lattice with
both absorbing and reflecting barriers\cite{Mur10,Kim25}. In general, 
the multifractal investigation for the chaotic and disorder systems
have been applied analytically and numerically to the transport phenomena
\cite{Has13,Bak14} and the random fractal structure having the time-dependent
random potential in the diffusive motion\cite{Bou15}-\cite{Fen18}.
Murphy et al\cite{Mur10} recently has argued on the 
multifractal behavior for the mean first passage time in one dimensional 
lattice having both absorbing and reflecting barriers.    
Therefore, we think that it is of great interest for the multifractal investigation 
to treat with 
the transport process described by the transition probability such as logistic map.\\ 
\indent
The purpose of this paper is to investigate the multifractal behavior of a particle executed 
on the random walk in the two dimensional Sierpinski gasket.
Specifically, we deal with the random walk for one particle 
having the transition probability given by the logistic map, and
concentrate on the  normalized first passage time for the random  walk 
with symmetric and asymmetric transition probabilities on Sierpinski
gasket existing both absorbing and reflecting barriers.
In section $2$, we present the transition probability expressed 
in terms of the logistic map, and 
the convenient formulas of the normalized first passage time and the multifractal are introduced. 
In section $3$, the distribution of normalized first passage time arriving at an absorbing
barrier is considered in both  Sinai model and the logistic model which has the
transition probability given by logistic map. For simplicity, the generalized
dimension  and the spectrum  from the statistical value of normalized  first passage time are
also calculated numerically,
and compared with the results 
reported earilier \cite{Kim25}.
Lastly, we give some conclusions. \\
\section{ Normalized first passage time and multifractals}
First of all,we present the transition probability as a chaotic orbits of logistic map
in this section. We also introduce the formulas of the normalized first passage time,
the generalized dimension, and the spectrum.\\
\indent  
Next, we shall consider the random walk of a particle in two-dimensional Sierpinski
gasket in which 
the number of sites $N_n$ having stage $n$ in the d-dimensional Sierpinski
gasket\cite{Ang19} is
given by $N_n= (d+1)(1+(d+1)^n)/2$.  It is assumed that a particle is started from an
arbitrary site $A$ on Sierpinski gasket. The reflecting barriers are located on
all the boundary, except that an absorbing barrier is at an ended site $B$ under
the right side, as shown in Fig.1.\\
\indent  
Now, the transition probability for the motion of a particle is introduced as 
\begin{equation}
x(n+1)  = Rx(n)[1-x(n)]
\label{eq:abc1}
\end{equation}                                
\\
where $R$ is the controll parameter with $0<R<4$.  In Monte Carlo simulation 
one has to treat with
the arbitrary values of $R$, where the numerical generated sequences appear to
be chatic\cite{Bak14,Sch20}.  By using the transition probability as Eq.~\ref{eq:abc1}, 
one particle jumps to right site with $p_{1j}+\gamma$ or left site with $q_{1j}-\gamma$ , 
and to up site with $p_{2j}+\gamma$    or down site with $q_{2j}-\gamma$ in four directions 
after a particle starts from a site $j$ on two-dimensional Sierpinski gasket, where $\gamma$ is 
called the disorder parameter.  It has really been showed that
the normalized condition for transition probability 
is given by $p_{1j}+q_{1j}+p_{2j}
+q_{2j}=1$. In particular, the transition probability
is symmetric  for the disorder parameter $\gamma=0$, and 
for $0<\gamma<\frac{1}{4}$, it also assumes that a particle executes on a biased random walk
jumped to the direction of the absorbing barrier.\\
\indent
Using the generating function technique the mean first passage time\cite{Sin2,Sin21} 
$<T>$ in one dimensional lattice is given by 
\begin{equation}
<T> = \sum_{k=1}^{N-1} \frac{1}{p_k} + \sum_{k=0}^{N-2} \frac{1}{p_k}  
\sum_{i=k+1}^{N-1} \prod_{j=k+1}^{i} \frac{q_j}{p_j}.
\label{eq:bcd2}
\end{equation}      
\\                 
where one particle moves to right site with $p_j$ or left site 
with $q_j$ after one step, when it starts from site $j$. However, the transition probability 
has always the different value at each site when a particle jumps for ocurrence of the 
one step to a nearest-neighbor
site from a given site, and the derivation for Eq.~\ref{eq:bcd2} can be found in details 
elsewhere\cite{Nos8,Mur81,Mur10}. 
Particularly, a possible extension of this paper is only to two dimension  
in order to discuss briefly on the mean first passage time and the multifractal in two 
dimensional Sierpinski gasket.
The normalized first passage time $T_{nt}$ has the form  
\begin{equation}
T_{nt} = \frac{t_n - t_{min}}{t_{max}-t_{min}} 
\label{eq:cde3}
\end{equation}
\\                       
where  $t_n$ is the first passage time arrivng at the absorbing barrier after $n$ steps, and
$t_{max}$ and $t_{min}$ are the maximum and minimum values of
the first passage time arrivng at the absorbing barrier, respectively. 
In the next section we will perform the conventional numerical simulations
for the normalized first passage time in both Sinai model and the logistic model
on two-dimensional Sierpinski gasket.\\ 
\indent
To show the multifractal feature for the motion of a particle with the
transition probability expressed in terms of the logistic function, we extend
to the multifractal calculations for the normalized first passage time.  If we
divide the normalized first passage time into a set $\epsilon$ of as $\epsilon \rightarrow 0$,
then the generalized dimension in the multifractal structure\cite{Has13,Bak14} is represented as
\begin{equation}
D_q = \lim_{\epsilon\rightarrow0}\frac{ln\sum_{i}n_{i}{p^q}_i}{(q-1)ln\epsilon} 
\label{eq:def4}
\end{equation}
\\                       
where $p_i$ is the probability for the number of configuration of $i^{th}$ particle
arrived at the absorbing barrier and $n_i$ the number of configuration for $i^{th}$ particle.
By introducing the above expression the spectrum  $f_q$ and $\alpha_q$ is caculated from the
relations
\begin{equation}
f_q = \alpha_q-(q-1)D_q 
\label{eq:efg5}
\end{equation}                             
\\
and
\begin{equation}
\alpha_q = \frac{d}{dq}[(q-1)D_q], 
\label{eq:fgh6}
\end{equation}                              
\\
where Eq.~\ref{eq:efg5} and Eq.~\ref{eq:fgh6} can be obtained by Legendre transformation. 
\section{Calculation and results}
We mainly concentrate on the generalized dimension and the spectrum for the
normalized first passage time on two-dimensional Sierpinski gasket lattice.
We only restrict ourselves to the case of $n=7$ stages
on Sierpinski gasket, even though the stage $n$
can be extended to large number in this paper.
At first, three sites are at $(1), (2),$ and $(3)$ 
in the initial stage $n=0$, and
we assume that a particle starts from an initial point $i_0 =(11111233)$ at $n=7$ stages
on two-dimensional Sierpinski gasket lattice. The reflecting barriers
is located on all the boundary surface, except that an absorbing barrier is 
at $B=(33333333)$ in Fig.$1$.
Next, both the cases of Sinai model and the logistic model, for asymmetric as well as
for symmetric transition probability, are considered as follows:
One is the case of Sinai model in which it has the symmetric transition
probability with $p_{1j}=q_{1j}=p_{2j}=q_{2j}=\frac{1}{4}$ and $\gamma=0$, and the two 
asymmetric transition probabilities with $\gamma=0.05$ and $\gamma=0.1$. 
The other is the case of the 
logistic model which is related to the transition probability given by 
the logistic map under the same condition of  Sinai model. \\
\indent
From now on, we estimate numerically 
the generalized dimension and the spectrum 
after finding the normalized first passage time from Eq.~\ref{eq:cde3} via
Monte Carlo simulation.
To discuss on these multifractal quantities for the
normalized first passage time in two dimensional Sierpinski gasket,
we perform the numerical simulations to show our analytic arguments. 
For three values of the disorder parameter (i.e., $\gamma=0, 0.05, 0.1$) 
in both Sinai model and the logistic model,
our simulations are performed $3\times10^6$ particles and averaged over $10^4$ configurations,
and analyzed the normalized first passage time, the generalized dimension, and the spectrum.
The result of these calculations is summarized in Table $1$. \\
\indent
Particularly, in the logistic model we are actually interested in the one case for the
control parameter $R=3.9999$. It can be easily found from the iteration of the logistic map
that three symmetric cutoff values are
$0.152210$, $0.505753$, and $0.855890$ for $\gamma=0$, while three asymmetric cutoff
values are $0.280024$, $0.505753$, and $0.730526$ for $\gamma=0.1$, respectively. 
Here the cutoff value is defined as the quantity used to determine the direction of
random walker in the logistic model. \\
\indent
As shown in Table $1$, the fractal
dimenion $D_o$( i.e., the maximum value of $D_q$ or $f_q$ ), the scaling exponent $\alpha_q$,
and the generalized dimension $D_q$,
ultimately based on the theoretical expressions 
Eq.~\ref{eq:def4} - Eq.~\ref{eq:fgh6}, 
are calculated numerically in both Sinai model and the logistic model. 
In Sinai model Figs. $2$ and $3$ depict respectively the generalized dimension
$D_q$ as a function of $q$ and the spectrum $f_q$
as a function of $\alpha_q$ for varying the disorder parameter values.
It can be seen from Table $1$ that the fractal dimension
changes anomalously as the disorder parameter $\gamma$ increases in two models.
In particular, 
it is also found from the result obtained in our simulations  
that in Sinai model for $\gamma=0.05$ the value of the fractal dimension on Sierpinski gasket 
is nearly equal to that on square lattice. 
Futhermore, in the logistic model on the two dimensional Sierpinski gasket,
the normalized first passage time is found to be infinite for $ \gamma =0$, and
the fractal dimension is expected to take the value near zero as $ \gamma\rightarrow 0.5$. \\
\indent
In conclusion, we have here considered the normalized first passage time
for both Sinai model and the logistic model 
on two dimensional Sierpinski gasket. 
We have clearly showed the multifractal
characteristics from the distribution of the normalized first passage time, 
as summarized in Table $1$. In future work, for the multifractal 
characteristics for the normalized first passage time,
we intend to compare with the results found in both Sinai model and the logistic model   
if the transition probability with the modified 
log - normal function is introduced, and to attempt to investigate extensively
in the similar lattice models. 
\section*{Acknowledgments}
This work was supported in part by Korean Dongwon Research Foundation.
\newpage
\section*{Figures}
\vspace{0.5cm}
Fig.1. The trajectory of a particle in the two dimensional Sierpi-
\indent \indent inski gasket at $n=7$ stages. A particle is started from at
\indent \indent $A=(11111233)$, and the absorbing barrier represents the
\indent \indent site $B=(33333333)$. \\ [0.5cm]
Fig.2. The generalized dimension  $D_q$ versus $q$ for the normal-
\indent \indent ized first passage time on the two dimensional Sierpinski
\indent \indent gasket in Sinai model. \\ [0.5cm]
Fig.3. The spectrum $f_q$ versus $\alpha_q$ for the normalized first pas-
\indent \indent sage time on the two dimensional Sierpinski gasket in Sinai
\indent \indent model. \\ [2cm]
\section*{Table}
\vspace{0.5cm}
Table I. Values of the fractal dimension, the generalized dimen- 
\indent \indent sion, and the scaling exponent in Sinai model and the logis- 
\indent \indent tic model. These lattice models are on both Sierpinski
\indent \indent gasket and the finite square lattice\cite{Kim25}.\\
\end{document}